\documentclass[aps,prl,twocolumn,showpacs,superscriptaddress,groupedaddress]{revtex4}
\usepackage{amsmath}
\usepackage{graphicx}
\usepackage{natbib}
\usepackage[usenames]{color}
\usepackage[T1]{fontenc}
\usepackage{textcomp}

\begin{document}

%\begin{frontmatter}

\title{Coexistence of cooperators and defectors in well mixed populations 
mediated by limiting resources}
% under limited resources}}
%and the evolutionary emergence of higher selective units.}}

\author{R.~J.~Requejo}
%\affiliation{Departament de F\'isica, Universitat Aut\`onoma de Barcelona, Campus UAB, E-08193 Bellaterra, Spain.}
\author{J.~Camacho}
\affiliation{Departament de F\'isica, Universitat Aut\`onoma de Barcelona, Campus UAB, E-08193 Bellaterra, Spain.}
\email {juan.camacho@uab.es}

\date{\today}% It is always \today, today,

\begin{abstract}
Traditionally, resource limitation in evolutionary game theory (EGT) is assumed just to impose a constant population size. Here we show that resource limitations may generate dynamical payoffs able to alter an original prisoner's dilemma (PD), and to allow for stable coexistence between unconditional cooperators and defectors in well-mixed populations. This is a consequence of a self-organizing process that turns the interaction payoff matrix into evolutionary neutral, and represents a resource-based control mechanism preventing the spread of defectors.
To our knowledge, this is the first example of coexistence in well-mixed populations with a game structure different from a snowdrift game. 
%Some implications on biological and economical contexts are outlined.
%These results present an alternative to the central control mechanism necessary for the emergence of higher selective units.
\end{abstract}


\maketitle

Cooperative behaviors are common in nature, and necessary for the evolutionary appearance of higher selective units --such as eukaryotic cells or multicellular life-- from simpler components 
%--prokaryotic cells, individual eukaryotic cells-- 
\cite{maynard-smith:1995a}. However, the survival of the fittest under the action of natural selection seems to foster selfish behaviors taking advantage of other individuals \cite{darwin:1859,dawkins:1989}. It is therefore intriguing how cooperative behaviors can emerge and survive in a world ruled by natural selection. 
This issue is frequently addressed in the context of evolutionary game theory (EGT), where the prisoner's dilemma (PD) game \cite{doebeli:2005} has been used as a paradigm for understanding the evolution of cooperation, as its simplified non-iterated version is the worst scenario for the survival of cooperation \cite{taylor:2007}; 
if interactions between individuals follow a PD and reproductive success grows with payoffs, cooperative behavior is led to extinction in large well-mixed populations \cite{taylor:2007,hofbauer:2003}. 
In the last decades some mechanisms have been found allowing cooperative behaviors to survive in the absence of genetic relatedness, but none of them works for the simplified PD in the absence of features such as memory \cite{trivers:1971,axelrod:1981}, reputation gain \cite{nowak:1998}, network structure \cite{hauert:2005, gomez-gardenes:2007, roca:2009} or other sensory inputs \cite{riolo:2001, traulsen:2003}. 

Here we study the influence of resource limitation on the emergence of cooperation. %Remarkably, 
We find that in addition to imposing a finite population size, as usually assumed in EGT \cite{doebeli:2005,taylor:2007,schuster:1983,trivers:1971,axelrod:1981,nowak:1998,riolo:2001, traulsen:2003,hauert:2005,gomez-gardenes:2007,roca:2009}, it may generate dynamic payoffs \cite{tomochi:2002,lee:2011,requejo:2011}.
In the absence of resource limitation the interactions between cooperators and defectors fulfil a simplified PD, as determined by the selfish strategy, 
and thus cooperators extinguish, as expected in evolving well-mixed populations. 
Surprisingly, resource limitation may drive a self-organizing process that allows for stable coexistence between cooperators and defectors. In contrast to previous studies including ecological features, in which coexistence happens only in public goods games with variable interaction group sizes \cite{hauert:2006}, it is transient  \cite{melbinger:2010} or requires spatial structure \cite{wakano:2007,dobramysl:2008}, here we find stable coexistence for pairwise interactions without population structure. This stable coexistence resembles the homeostatic equilibrium in the daisy world \cite{lovelock:1974,watson:1983}, as both are mediated by environmental factors driving the system out from equilibrium.
%Here, however, we consider direct interactions between individuals, which paves the way for the emergence of new units of selection, such as eukaryotes or multicellular organisms, from an initial mixed state of cooperative and non-cooperative individuals. 

In order to study this, we develop a model consisting of an evolving well-mixed population of self-replicating individuals that receive resources from the environment and exchange resources during interactions. No population structure, memory, learning abilities or any other sensory inputs are assumed. Each individual $i$ is represented by its internal amount of resources $E_i$ and its strategy, namely cooperate (C) or defect (D). The internal amount of resources may be interpreted as the amount that belongs to it, independently of why or how. The environment provides resources in portions $E_{0}^{i}$ per unit time to randomly chosen individuals independently of their strategy, thus not modifying the structure of the payoffs. For simplicity, we impose a constant total resource influx $E_T$, though results also apply for non constant fluxes (see \cite{sm:prl2011}). If the amount of resources of an individual exceeds a value $E_s$, it splits into two identical copies with half its internal amount of resources.

Defectors are characterized by the maximum amount of resources associated to an interaction: the cost spent $(E_c)$ for stealing a reward $(E_r)$ from the co-player. If the internal resources of a defector are smaller than the cost $E_c$, it does not pay the cost nor receives the reward. If the interaction partner has resources below the reward, the entire amount of resources is sequestered. We assume that these quantities are inherited without mutation; they represent physiologic, morphologic or genetic characteristics intrinsic to individuals and cannot be modified by choice. 

We consider large populations, simultaneous interactions and $E_r>E_c>0$, although the same results are obtained if one assumes that every interaction is carried out by a donor and received by the co-player \cite{requejo:2011}. The interaction matrix determined by the strategies is thus
\begin{equation}
\label{(2.2)}
\bordermatrix{\text{}&$C$&$D$&\cr
                $C$&0 & - E_r\cr
                $D$& \Delta E & - E_c\cr}
\end{equation}
and equals a simplified PD, with defectors paying a cost $E_c$ and obtaining a net reward $\Delta E=E_r-E_c>0$ (payoffs for the row player, we will omit C and D in the following matrixes). 

Finally, we assume that the limiting resource necessary for reproduction provides no advantage for keeping alive; therefore deaths occur at random with a frequency (rate) $f$ relative to receiving resources and interacting, which happen equally frequently. 

\begin{figure}
\includegraphics*[width=8cm]{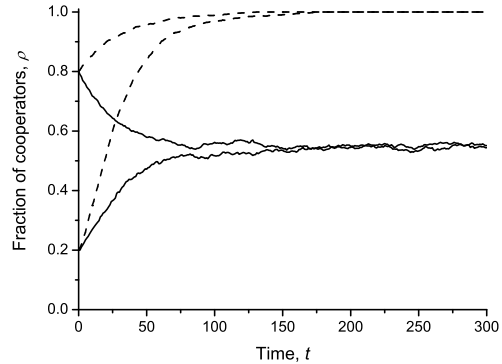}
\caption{\label{fig1} Simulation results for the evolution of the fraction of cooperators $\rho $ for two different values of the reward $E_r$ and cost $E_c$ associated to the selfish strategy (averaged over 10 runs). In some cases the simplified PD payoff structure is modified by the 
%existence of a limiting resource
limitation of resources, which allows for coexistence of cooperation and defection (solid line) and dominance of cooperation (dashed line). The final stable states are independent of the initial $\rho $ and $N$. Parameters: $f=1$,  $E_s=1000$, $E_T=8200000$, $E_c=660$; solid line, $\Delta E=310$; dashed line, $\Delta E=10$.
%; mean population sizes $\overline{N}=10450$ individuals (solid line) and $\overline{N}=11500$ individuals (dashed line)
 }
\end{figure}

According to the PD structure of resource exchanges among cooperators and defectors in the absence of resources limitation, defectors should have a larger resource intake, reaching faster the splitting bound $E_s$ and thus reproducing quicker (i.e. fitness is proportional to resource exchanges). Therefore one would expect homogeneous populations of defectors as the outcome of the evolutionary process. However, the limitation of resources generates a distribution of resources among individuals in the population. As a consequence, the average reward stolen from cooperators $E'_r$ may decrease below $E_r$, since the internal resources of some cooperators may fall below this quantity, thus modifying the payoffs. If the net average benefit of defectors $\Delta E' = E'_r-E_c$ remains negative all over the time, the payoff matrix does not correspond to a PD anymore; it is turned into a harmony game and cooperation becomes the dominant strategy. Simulations show that the model yields the later behavior and also, more interestingly, stable coexistence of cooperation and defection (Fig.~\ref{fig1}); see 
\cite{sm:prl2011} for details on the simulations. Dominance of cooperation was already found in a previous model assuming that resources are necessary for keeping alive \cite{requejo:2011} which, however, did not provide coexistence.

%{\color{red} This coexistence behavior does not occur if, in contrast to what considered here, one assumes that resources are also necessary for keeping alive \cite{requejo:2011}}.

Coexistence in this scenario requires a complex feedback process whose exact analysis is quite difficult because of the complex nonlinearities involved in the dynamics. However, a simple quantitative reasoning exhibits the logic of this feedback and allows for an analytic estimation of the final stable state of the system.
Let us note that an increase in the number of defectors over the equilibrium value would cause an overexploitation of cooperators, thus reducing their resource content. This would have two effects: (i) it would reduce cooperators' reproduction rate (fitness) because they become farther from the splitting bound $E_s$, and (ii) it  would also decrease the average reward obtained by defectors, which thereby reduces their fitness. If the second effect dominates over the first one, then stable coexistence becomes possible, as the feedback pushes the system back to equilibrium. A similar argument applies for a decrease in the number of defectors.

The entire system is in equilibrium when the resource influxes and out fluxes in the populations of both cooperators and defectors cancel out. The balance of resources in these subpopulations contains three contributions: environmental supply, 
deaths, and interactions. They are expressed in the following equations
\begin{eqnarray}
\label{(4.1a)}
\frac{dE^C}{dt} = N_C[E_0-f\overline{E}^C-pE'_r(1-\rho)] \\
\label{(4.1b)}
\frac{dE^D}{dt} = N_D[E_0-f\overline{E}^D-pE_c+pE'_r\rho]
\end{eqnarray}
$E^j$, $\overline{E}^j$ and $N_j$ denote, respectively the total resource content, average resources per individual and number of individuals of the subpopulations $j=C,D$ 
%(C stands for cooperators and D for defectors)
; $E_0=E_T/N$ is the mean amount of resources received by an individual per unit time, with $N=N_C+N_D$ the instantaneous population size; $\rho=N_C/N$ is the fraction of cooperators; $f$ is the death probability per individual and interaction, and $p$ the fraction of the population of defectors able to pay the cost (i.e. with $E_i>E_c$).

\begin{figure*}
\begin{center}
\includegraphics*[width=17cm]{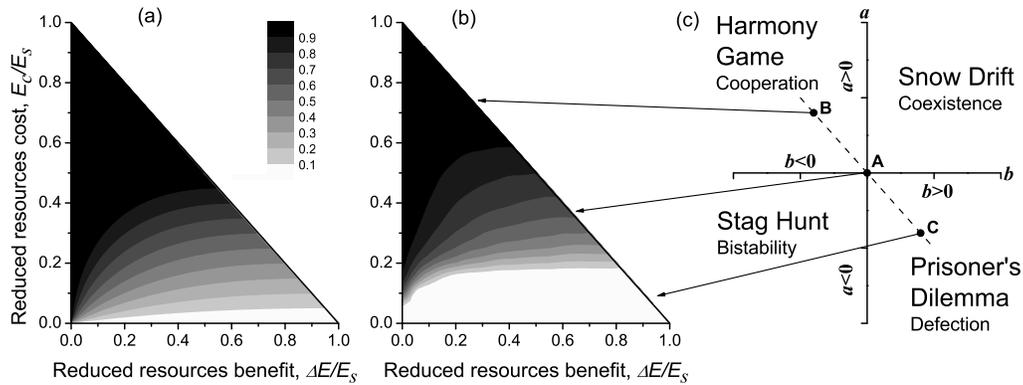}
\caption{\label{fig2}Final fraction of cooperators $\rho $ represented in terms of resources cost ($E_c$) and net benefit ($\Delta E = E_r - E_c$) of the selfish strategy. In black $\rho=1$, in white $\rho=0$. One observes well defined regions of coexistence of cooperation and defection, as well as regions where cooperation is the dominant strategy. In (a) prediction according to Eq.~(\ref{(4.5)}) (see \cite{sm:prl2011} for a highly improved analytical prediction); in (b) results of agent-based simulations averaged over 50 runs ($f=0.01$,$E_s=1000$, $E_T=420000$). In (c) we show the different games corresponding to a 2x2 matrix; the dashed line shows the places where the payoffs in the model lay (see Eq.~(\ref{(4.8)})). Point A denotes the final payoffs for coexistence states, where the payoff matrix is evolutionary neutral; points B and C are examples of final payoffs for situations where cooperation and defection are dominant, respectively.}     
\end{center}
\end{figure*}

In equilibrium, the populations of cooperators and defectors become constant in time so that the resource pools $E^D$ and $E^C$ reach a constant value. One thus finds the equilibrium condition
\begin{eqnarray}
\label{(4.2)}
p(E'_r-E_c) = f[\overline{E}^D-\overline{E}^C]
\end{eqnarray}
This shows that the coexistence depends on the death frequency $f$. For simplicity we will assume in this analytic derivation that deaths happen much less frequently than interactions, i.e. the limit $f \to 0$; this corresponds to many interactions in a lifetime, when the effects of interactions become more relevant. Other $f$ values are studied through simulations \cite{sm:prl2011}. Since $p$ never equals zero due to the constant resource influx, Eq.~(\ref{(4.2)}) reduces in this limit to
\begin{eqnarray}
\label{(4.3)}
E_c=E'_r
\end{eqnarray}
which states that, in equilibrium, the cost paid by defectors equals the reward stolen from cooperators. In order to analytically predict the region of coexistence in the parameters space and the corresponding population composition, we need to know the average reward $E'_r$ in terms of the parameters $E_r$ and $E_c$. This implies the calculation of the equilibrium distribution of resources for cooperators, which is a difficult task due to the nonlinearities involved in the dynamics. Instead, we can give a rough heuristic estimate as follows. The lower the fraction of cooperators in the population, the more frequent any cooperator meets a defector, thereby cooperators become overexploited and their average internal resources decrease. Thus the average reward $E'_r$ is expected to decrease as $\rho$ decreases. We assume a linear relationship between both quantities, $E'_r = \alpha \rho$, with $\alpha$ a positive constant. By the moment we consider that when $\rho$ is close to 1, the effect of defectors is expected to be small, so that at first order we approximate the resource distribution of cooperators as uniform. For uniform distributions \cite{sm:prl2011} one finds $E'_r = E_r - E^2_r/(2E_s)$. We thus propose
\begin{eqnarray}
\label{(4.4)}
E'_r = \rho (E_r- \frac{E^2_r}{2E_s})
\end{eqnarray}
By combining Eq.~(\ref{(4.4)}) with Eq.~(\ref{(4.3)}) one obtains an expression for the equilibrium fraction of cooperators 
\begin{eqnarray}
\label{(4.5)}
\rho = \frac{E_c}{E_r-E^2_r/2E_s}.
\end{eqnarray}

In order to analyze in detail the behavior of the model, we have performed extensive numerical simulations covering the whole parameters space. They confirm the stability of the coexistence for all death frequencies, and show that the final stable state is independent of the initial conditions and resource influx (and thus, final population size) \cite{sm:prl2011}.
Figs.~\ref{fig2}a,b show a good qualitative agreement between the predicted $\rho$ 
%Eq.~(\ref{(4.5)}).
and the outcome of the simulations. Deviations root in the linear approximation assumed in Eq.~(\ref{(4.4)}) (see \cite{sm:prl2011}).

Let us notice that the obtained stable coexistence between cooperators and defectors presents a new outcome in the context of two-player games, where a stable mixed state is only expected in Snowdrift (or Hawk-Dove) games, which have a payoff structure different from ours. In general, symmetric two-player games can be described through the interaction matrix \cite{sigmund:2010} 
\begin{equation}
\label{(4.6)}
\begin{pmatrix}
0 & a \\[4mm]
b & 0
\end{pmatrix}
\end{equation}
where coefficients $a$ and $b$ are assumed to be constant. Applying the replicator equation \cite{schuster:1983,hofbauer:2003,sigmund:2010} to analyze the evolution of the population, three cases are possible (see Fig.~\ref{fig2}c): (i) dominance of one strategy (when $a$ and $b$ differ in sign); this is the case of the non-iterated PD, where defection always wins; (ii) bistability (if both $a$ and $b$ are negative), in this case the final state is homogeneous and depends on initial conditions; this is what happens in stag hunt games, where coordinating with the partner pays; and (iii) coexistence (if both $a$ and $b$ are positive); this is what occurs in Snowdrift games, when it always pays to play the opposite of the co-player. 

In our model, fitness is directly proportional to resource exchanges, because individuals reproduce when their resources overcome an upper bound that is the same for cooperators and defectors. Resource exchanges come from the environment and from interactions. The resource supply from the environment is the same for defectors and cooperators; it just provides a constant to all fitness values and can be omitted in the fitness matrix. The latter is thus ruled by the average resources exchanged through interactions, which aside from a scale factor translating resource exchanges to fitness, is \cite{requejo:2011}
\begin{equation}
\label{(4.7)}
\begin{pmatrix}
0 & -pE'_r \\[4mm]
p\Delta E' & -pE_c
\end{pmatrix}.
\end{equation}
As stated above, $p$ stands for the fraction of defectors whose resources exceed the cost $E_c$. Let us note that this factor does not change the payoff structure in any case, as it multiplies all payoffs, and it only modifies the time scale of the dynamics. The interaction matrix can be rewritten in the form of matrix (\ref{(4.6)}) by adding $pE_c$ to the second column (as adding a constant to a column does not affect the replicator dynamics \cite{hofbauer:2003,sigmund:2010}): 
\begin{equation}
\label{(4.8)}
\begin{pmatrix}
0 & -p\Delta E' \\[4mm]
p\Delta E' & 0
\end{pmatrix}
\end{equation}
i.e. $a=-b=-p\Delta E'$. According to the classification given above, this payoff matrix leads to dominance of one strategy whenever $p\Delta E'\neq 0$. 
In the absence of resource limitation $\Delta E'=\Delta E>0$ and we have a PD. If resources are limited, there exists a wide range of parameters for which the $\Delta E'$ is tuned to zero for a specific mixture of cooperators and defectors (see Figs.~\ref{fig2}a,b); thus, the stable equilibrium is the result of a dynamical self-organizing process and not of the game structure itself  (see Fig.~\ref{fig2}c). 

We can use the payoff matrix (\ref{(4.8)}) to gain further insight into the stability of the coexistence state found in our model. In Eq.~(\ref{(4.4)}) we proposed the rough estimate $E'_r=\alpha \rho $ for the net benefit of defectors, with $\alpha>0$. Thus, we have $\Delta E' = \alpha \rho - E_c$. Aside from a positive factor relating fitness and payoffs in Eq.~(\ref{(4.8)}), the replicator equation yields
\begin{eqnarray}
\label{(4.9)}
%d\rho /dt = -\rho (1-\rho ) p \Delta E = p \rho (1-\rho )(E_c- \alphaup \rho )
\frac{d\rho}{dt} = -\rho (1-\rho )p\Delta E = p\rho (1-\rho )(E_c-\alpha \rho )
\end{eqnarray}
which supplies three equilibria, $\rho = 0, 1$ and $E_c/ \alpha $.
Since $p>0$, the mixed state is the stable one for $0<E_c/ \alpha <1$, in agreement with the stability of the coexistence states observed in the simulations. 

We have presented a scenario which allows for stable coexistence of unconditional cooperators and defectors in well-mixed populations under pairwise interactions. This result is quite robust, since it does not depend on initial conditions, and it is also observed in small populations -- though in this case fluctuations may lead to the extinction of one strategy -- and under nonconstant influx of resources (\cite{sm:prl2011}). This stable coexistence roots on a selforganizing process which implicitly includes the environment, and it is the feedback induced by environmental constraints and defectors' behavior which turns the payoff matrix into evolutionary neutral and allows for the stability of the system. The evolutionary neutrality of the system (environment + individuals) and its stability as a whole, might be a first step towards the emergence of new units of selection by providing a selforganizing mechanism preventing the spread of selfish mutants alternative to central control (see \cite{maynard-smith:1995a}).
%beyond the usual assumption of full cooperation.
%In particular, the robustness of our results, which allow for long term stability, and the simplicity of the effect described here, which does not need any evolved sensory input, suggests that it might have played a role in the endosymbiotic appearance of eukaryotic cells and the transition from unicellular to multicellular organisms.

Let us also remark that, in contrast to previous models in evolutionary dynamics, the model presented here explicitly sets the issue in a nonequilibrium context, where a (resource) flux drives the system out from equilibrium. The observed selforganized coexistence state may be seen as another example of selforganizing process found in nonequilibrium systems such as, for instance, the unexpected oscillations in Belusov- Zabhotinsky reactions. This perspective may bear interest in economic contexts, another classical field of evolutionary game theory, where some authors claim that economic systems should be modeled as open, nonlinear nonequilibrium systems instead of the closed, equilibrium view dominant in traditional economics \cite{beinhocker:2006,peters:2010}.
%, and where classical models used in economics display a much richer behaviour than expected under nonequilibrium settings \cite{Tauber}.  

We thank D. Jou, X. Alvarez and D. Bassignana for useful discussions. This work has been supported by the Spanish government (FIS2009-13370-C02-01) and the Generalitat de Catalunya (2009SGR0164). R.J.R. acknowledges the financial support of the Universitat Autònoma de Barcelona and the Spanish government (FPU grant).

\end{document}

%\begin{frontmatter}

\title{Supplementary material of ``Coexistence of cooperators and defectors in well mixed populations mediated by limiting resources.''}

\author{R.~J.~Requejo}
\affiliation{Departament de F\'isica, Universitat Aut\`onoma de Barcelona, Campus UAB, E-08193 Bellaterra, Spain.}
\author{J.~Camacho}
\affiliation{Departament de F\'isica, Universitat Aut\`onoma de Barcelona, Campus UAB, E-08193 Bellaterra, Spain.}
\email {juan.camacho@uab.es}


\begin{abstract}
In this supplementary material we provide details of the analytical derivations carried out in the main text and of the agent based simulations.
\end{abstract}

%\pacs{}

\maketitle

\section{Simplified prisoner's dilemma (PD) game.}
 The rank of the payoffs in the PD is 
\begin{equation}
T>R>P>S.
\end{equation} 
If the game is played repeatedly by the same two individuals (iterated PD), which happens with a very low probability as we assumed big population sizes (in order to make mean field approximation), the condition $2R>T+S$ is also required. 
For a simplified PD, the condition $T-R=P-S$, also known as equal gains from switching, must be fulfilled. In our study, the payoffs take values $T=E_r-E_c$, $R=0$, $P=-E_c$, $S=-E_r$ and the three conditions are fulfilled by imposing 
\begin{equation}
E_r>E_c>0.
\end{equation} 

\section{Resource distribution of cooperators: estimation of $E'_r$.}

\subsection{Uniform distribution of resources}
Let us call $P(E<E_r)$ the probability that a cooperator has an internal amount of resources lower than $E_r$. The mean payoff for a defector playing against a cooperator can be written as $E'_r = P(E>E_r)E_r + P(E<E_r)\overline{E}_r$, where $\overline{E}_r$ is the mean internal amount of resources of cooperators in the region $E<E_r$. This may be rewritten as $E'_r = E_r - P(E<E_r) (E_r- \overline{E}_r)$. For uniform distributions of resources this equation yields 
\begin{equation}
E'_r = E_r-\frac{E^2_r}{2E_s}.
\end{equation}

\begin{figure}
\includegraphics*[width=8cm]{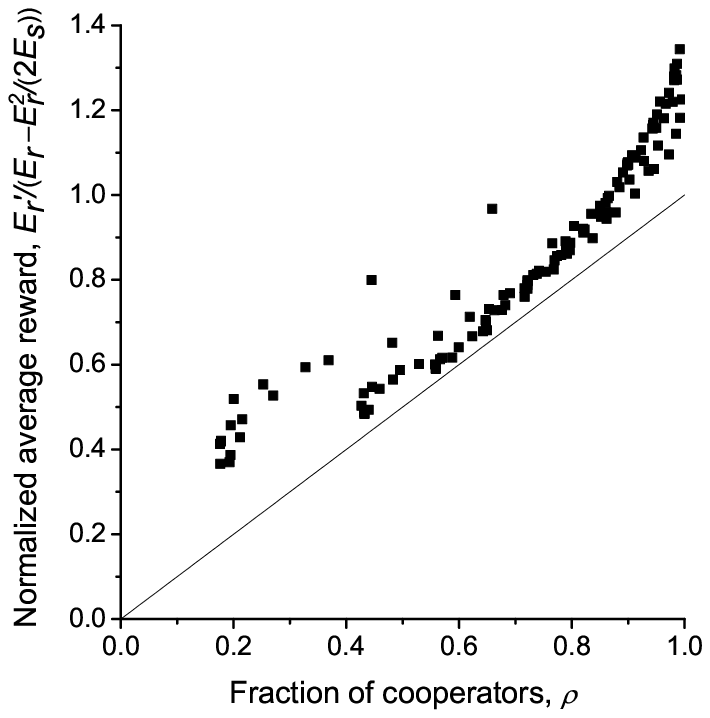}
\caption{\label{figsm:1} Normalized average reward of defectors $E'_r/(E_r-E^2_r/(2 E_s)$ versus fraction of cooperators in equilibrium $\rho$. Symbols denote simulation results expanding over the whole parameter space for $f=0.01$. The straight line corresponds to the assumption of a linear relationship between $E'_r$ and $\rho$. One finds a reasonable agreement between both results with some deviations at low and high $\rho$.}
\end{figure}

\subsection{Linear approximation for $E'_r$}
\begin{figure}
\includegraphics*[width=9cm]{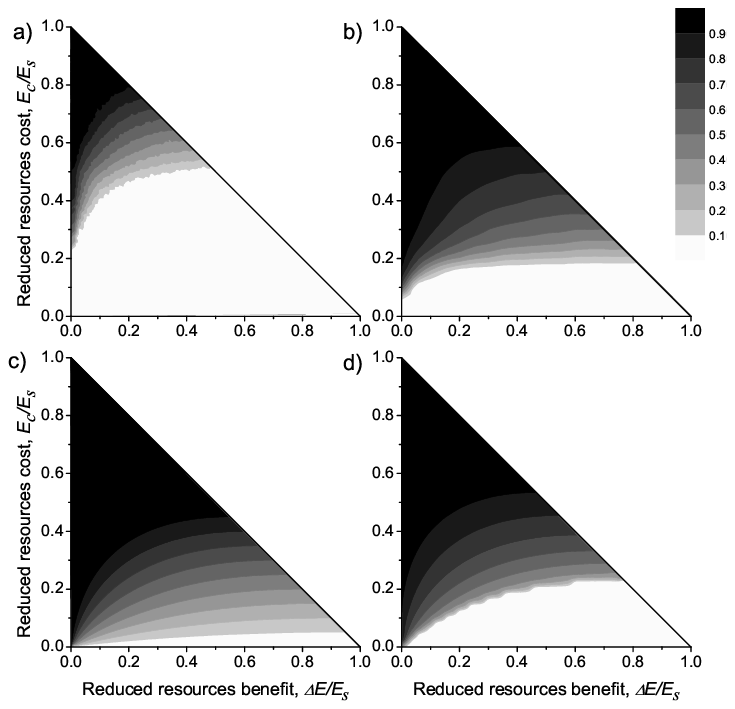}
\caption{\label{figsm:2} Simulation results for (a) $f=1$, i.e. interactions and deaths equally frequent, and (b) $f=0.01$, i.e. a hundred interactions per death. The regions of coexistence and dominance of cooperation grow with decreasing $f$. Analytic results for $f=0.01$  assuming (c) a linear dependence, and (d) a quadratic fit between $E'_r$ and $\rho$ as obtained from figure 1 above. The analytic predictions improve by using the quadratic fit, as it can be observed comparing figures (b) and (d), but the simple linear relationship already gathers the main features.}
\end{figure}

To derive Eq.(7) in the main text we assumed the linear dependence $E'_r = \alpha \rho$, where $\alpha$ is chosen to attain expression (3) when $\rho=1$.
Fig.~\ref{figsm:1} shows a reasonable agreement between this simple expression and the simulation results. As assumed, the average reward $E'_r$ is found to grow with the fraction of cooperators, although a quadratic dependence is more adequate than just a linear one. If one assumes such a quadratic dependence the agreement between predictions and simulations greatly improves (see Figs.~\ref{figsm:2}b,d). 

\section{Numerical simulations.} 

The final state of the system was found to be independent of the initial conditions, i.e. initial population size $N_0$, fraction of cooperators $\rho$ and initial distritution of resources, provided that the initial population size is big enough (see section \ref{sec:ci} below). In Figs. 2 both in the paper and the SM, we started all simulations with a population size $N_0 = 10000$, a uniform random distribution of internal resources and an initial fraction of cooperators of $\rho=0.5$. The amount of resources for splitting was taken $E_s=1000$, which sets the scale of resources in the simulations. For $E_T$ we used 420000, so that the total population $N$ attained values of several thousands, large enough to minimize finite size effects; however, the mechanism works for much smaller population sizes (smaller than 100 individuals, see fig. \ref{figsm:5}), provided that random fluctuations on the number of individuals do not allow the system to reach fixation. Simulations run over around 1000 time steps, where a time step is defined as a number of interactions equal to the population size, and stopped if a homogeneous population was reached before. 

\subsection{Resource allocation}
We tested two resource allocation methods: supplying portions uniformly distributed on the interval $E_{0}^{i}\in(0, 2E_T/N)$ and of fixed size $E_{0}^{i}=E_T/N$, where $N$ denotes the number of individuals in the population; both provided the same results, but the first one (which is the one used in the figures) introduces some extra stochasticity in the model. As said above, $E_T$ was chosen big enough as to avoid effects due to finite population sizes, while keeping feasible simulation times (mean population sizes around $10^4$ individuals).

\subsection{Updating}
The updating is completely asynchronous \cite{szabo:2007} and with overlapping generations, which avoids the appearance of synchronization effects and mimics reproductive dynamics observed in nature. The implementation used is as follows: Every interaction time step six individuals are chosen at random; of them (a) two receive an amount of resources $E_{0}^{i}$ from the environment, (b) two interact, and (c) two die with a probability $f$. Other asynchronous implementations where tested, as choosing just four individuals per interaction step, one to act as donor, one as recipient of the act, one to receive resources and one to die with probability $f$. Results using this updating were identical, but simulation times were much longer, for which reason we choose the first method to carry out the computer simulations.

\subsection{Remarks on the parameter space.} 
Our model contains 5 parameters: $E_T$, $E_s$, $E_c$, $E_r$ and $f$. If one doubles all the parameters related to the resources, i.e. all of them except $f$, the dynamics does not change, provided there are no finite population size effects. Then, one of the parameters in the model just defines the scale with respect to the others. We thus set $E_s=1000$ without loss of generality. Furthermore, as long as $E_T$ takes a large value ($E_T\gg E_s$), it only affects the final number of individuals, but not the strategy of the final state (this is confirmed by simulations). This leaves us with the free parameters, $E_c$, $E_r$ (both smaller than $E_s$) and $f$. We have explored the dynamics for all values of $E_c$ and $E_r$ in the region $E_s>E_r>E_c$, and for several values of $f$ (see Fig.~\ref{figsm:2}). In this way, our analysis covers the whole parameter space.

\subsection{Independence of results on initial conditions and on resource influx}
\label{sec:ci} 
\begin{figure*}
\includegraphics[width=14cm]{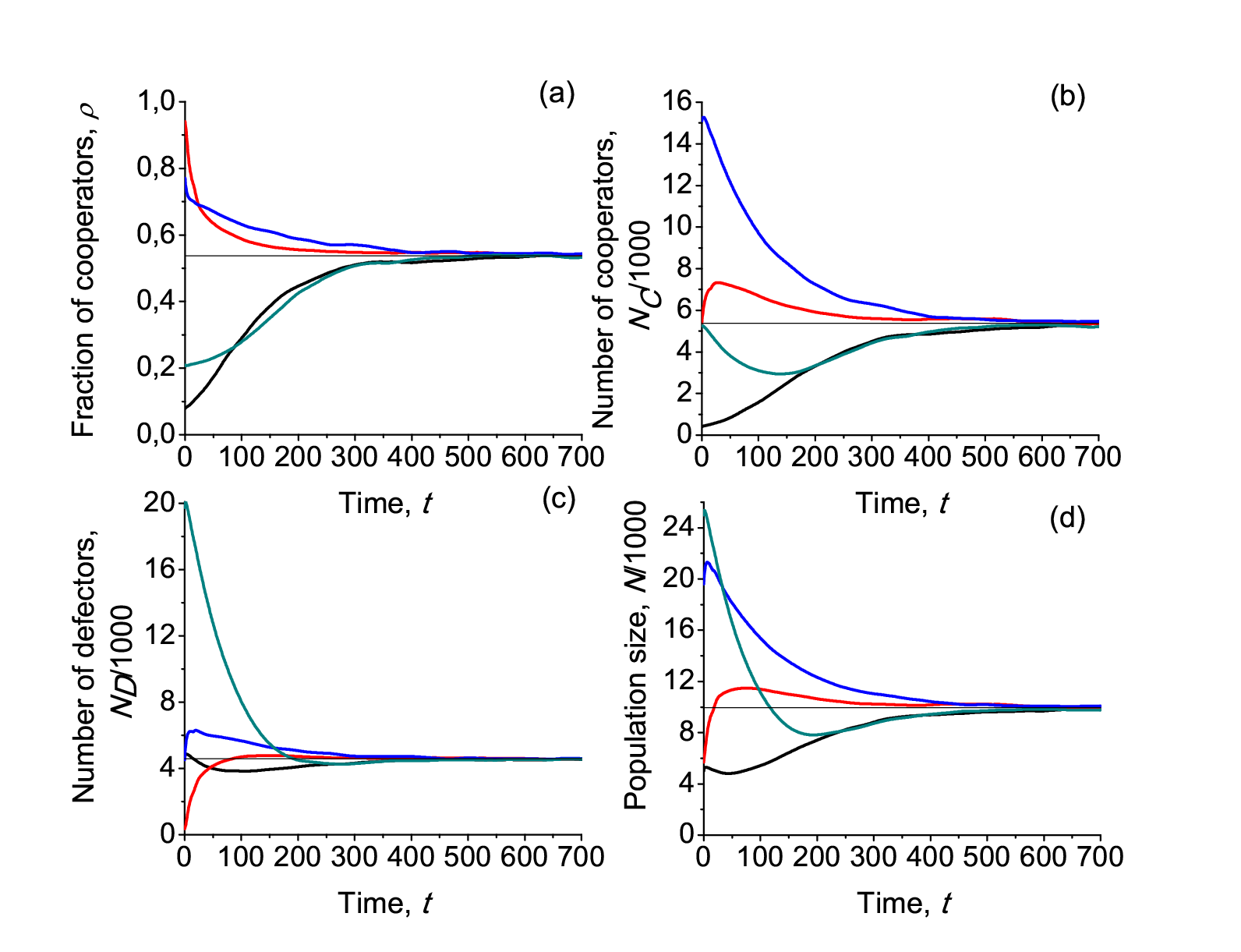}
\caption{\label{figsm:3} {Evolution of (a) $\rho$, (b) $N_C$, (c) $N_D$  and (d) $N$, for four different initial numbers of cooperators and defectors. It can be observed that in all cases, even when the initial values are far away from the final state, the system recovers its equilibrium values given there is enough time. In the case in which the initial number of defectors is much bigger than its final stable value (five times bigger, green lines), or that of cooperators is very small (13 times smaller, black lines), the population of cooperators/defectors (respectively) decreases initially, and thus this situation could drive one of the populations to extinction if its initial number was small enough (see Fig. \ref{figsm:5}). Parameters: $E_T = 420000, E_c=300, \Delta E=400, f=0.01$. Final values: $\rho = 0.54, N_C = 5380 , N_D = 4580, N = 9960$ }}
\end{figure*}

As mentioned above, the simulation results are very robust. In particular, they are independent on initial conditions aside from finite size effects when either $N$ or $\rho$ are too small. For instance, in figure \ref{figsm:3} we show simulations results starting from four different initial amounts of cooperators and defectors ($N_C,N_D$), all leading to the same final steady state. 

\begin{figure*}
\includegraphics[width=12cm]{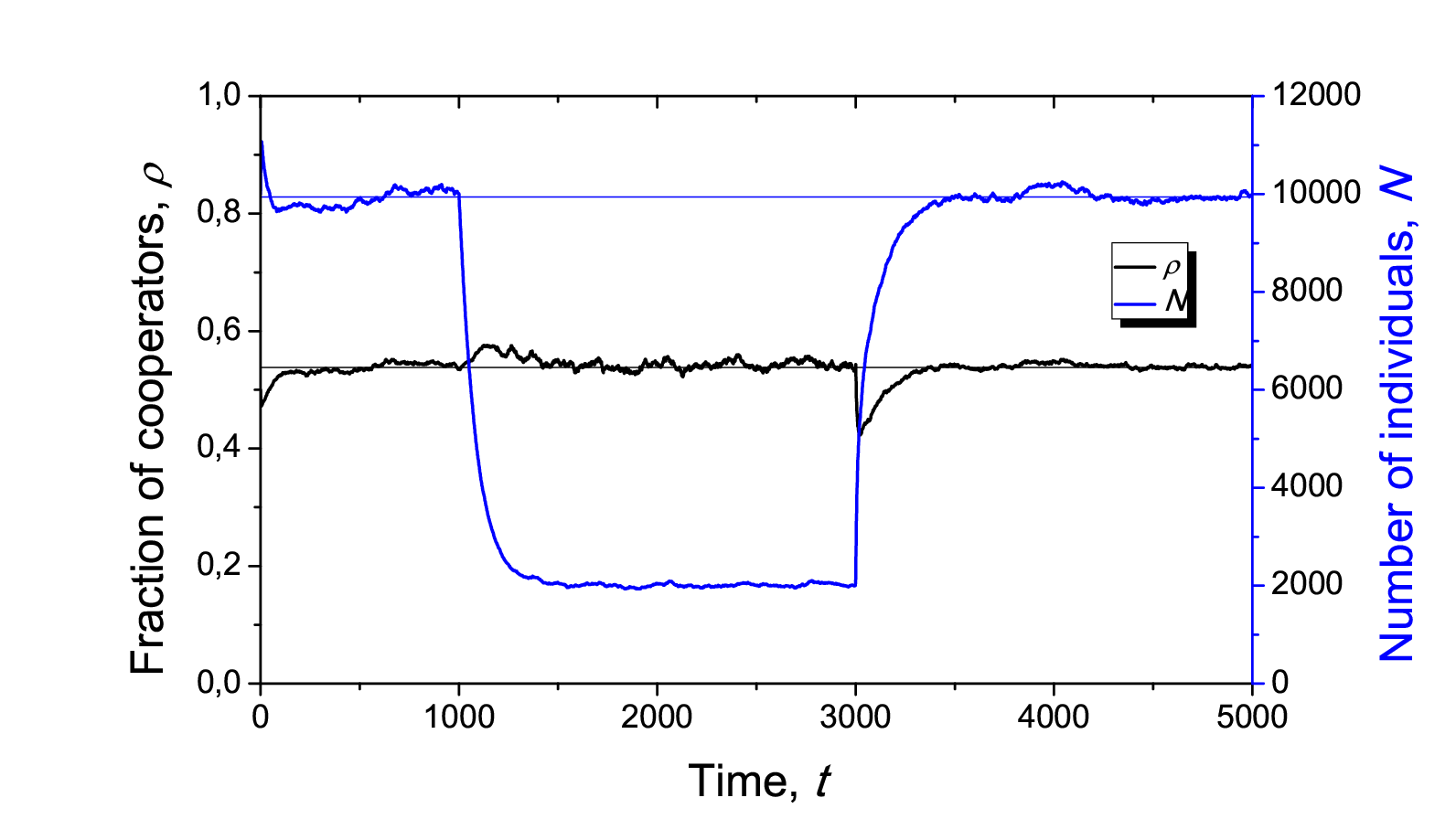}
\caption{\label{figsm:4} Simulation results dividing by five the total amount of resources $E_T$ introduced into the system  at $t=1000$, and recovering the initial value at $t=3000$. It can be observed that the population size behaves as $E_T$, decreasing in a factor 5 during the intermediate period, while the fraction of cooperators is only slightly affected during the short time spans in which the population increases or decreases. Parameters: $E_{T,inic}=420000, E_c=300, \Delta E=400, f=0.01$; values in equilibrium: $\rho = 0.54$, $N = 9960$}
\end{figure*}

In addition, our results are robust with respect to changes in the resource influx and, in particular, the latter is not required to be constant. Figure \ref{figsm:4} shows simulation results for a sudden reduction of $E_T$ to its fifth. As expected, the population size $N$ also reduces to its fifth, but the fraction of cooperators hardly changes. Indeed, it is observed a slight transient advantage for cooperators when resources are reduced, and a transient advantage of defectors when resources are increased. 

\begin{figure*}
\includegraphics[width=17cm]{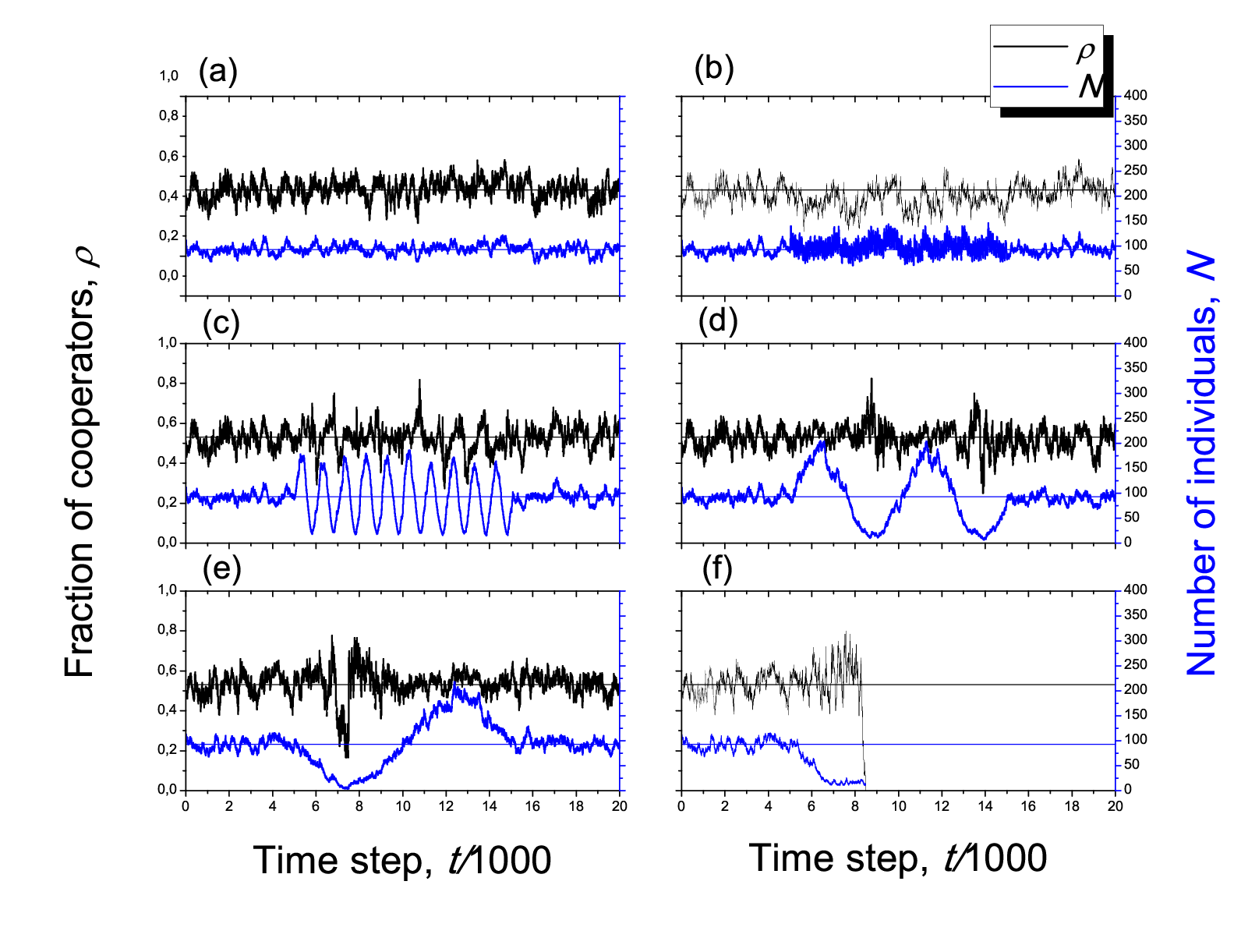}
\caption{\label{figsm:5} {Dynamics of small populations. Simulation results for (a) a constant $E_{T}=4000$, and a variable $E_{T}=4000+3500 sin(2\pi t/w)$ in between $t=5000$ and $t=15000$ with periods (b) $w=100$, (c) $w=1000$, (d) $w=5000$, (e),(f) $w=10000$}. It can be observed in figures (b)-(e) that the inclusion of a variable resource influx only introduces noise in the value of $\rho$, while $N$ follows the behaviour of $E_T$. For values of $w$ lower than 100 (the one in fig. (b)), $N$ does not follow $E_T$ and only some noise is introduced into the sytem. In (f) we show an example of fixation of one strategy: a random fluctuation makes the system remain in a state with 15 individuals during enough time as to allow for fixation of defectors. Note that typical fluctuations in the equilibrium (fig.(a)) have a standard deviation on $N_C$ of 8 individuals, while it is of 4 individuals for $N_D$; thus, any state in which the number of individuals is maintaind in very low values during time enough will reach fixation, being fixation of defectors favoured over coopeartors in this case. Mean values in equilibrium (fig.(a)): $\rho = 0.53$, $N = 93$. Parameters: $E_c=300, \Delta E=400, f=0.01$.}
\end{figure*}

Finally, Figure \ref{figsm:5}a analizes the limit of small population sizes. One observes that both the number of individuals in the population and the fraction of cooperators reach steady states with larger random flutuations. The other panels of Figure \ref{figsm:5} study the effect of oscillatory changes of the total resource influx $E_T$ in these small populations. The population numbers are found to follow the resource influx $E_T$ (except at fast oscillations of period $w=10$, shorter than the time scale of the population dynamics). Remarkably, the fraction of cooperators remains essentially constant, though fluctuations may drive the system to the extinction of one strategy (Fig. \ref{figsm:5}f).

